\documentstyle[12pt]{article}
\newcommand{\be}{\begin{equation}}
\newcommand{\ee}{\end{equation}}
\newcommand{\ba}{\begin{eqnarray}}
\newcommand{\ea}{\end{eqnarray}}
\newcommand{\no}{\nonumber \\}

\begin{document}

\begin{titlepage}

\begin{center}
\LARGE\bf Fixed-Point Analysis of the Low-Energy Constants
in the Pion-Nucleon Chiral Lagrangian \\  

\end{center}
\vspace{1cm plus 0.5cm minus 0.5cm}
\begin{center}
{\large 

 Youngman Kim$^{(a, b)}$, Fred Myhrer$^{(b)}$ and  
 Kuniharu Kubodera$^{(b)}$}
\end{center}
\vspace{0.5cm plus 0.5cm minus 0.5cm}
\begin{center}

(a)~{\it  School of Physics, Seoul National University,
Seoul 151-742, Korea}\\
(b)~{\it Department of Physics and Astronomy, 
University of South Carolina, 
Columbia, South 
Carolina 29208}

\end{center}
\vspace{1.0cm plus 0.5cm minus 0.5cm}

\begin{abstract}
In the framework of heavy-baryon chiral perturbation theory, 
we investigate the fixed point structure 
of renormalization group equations (RGE) 
for the ratios of the renormalized 
low energy constants (LECs) that feature 
in the pion-nucleon chiral Lagrangian.
The ratios of the LECs deduced from our RGE analysis
are found to be in semi-quantitative agreement with 
those obtained from direct fit to the experimental data.
The naturalness of this agreement
is discussed using a simple dimensional analysis 
combined with Wilsonian RGEs.

\end{abstract}
\end{titlepage}

\newpage

\renewcommand{\thefootnote}{\#\arabic{footnote}}

\section{Introduction}
As is well known, 
the long-distance or low-energy behaviour of QCD
is not amenable to perturbative approaches.
Chiral perturbation theory (ChPT) provides
a highly useful framework for correlating
hadronic observables 
in the low-energy regime~\cite{chpta}.
In ChPT, possible terms 
in the effective Lagrangian, ${\cal L}_{\rm eff}$,
are classified in terms of the chiral index $\nu$;
the contribution of a term with index $\nu$
carries the factor $(Q/\Lambda)^\nu$,
where $Q$ is a typical energy-momentum scale
involved in a given process,
and $\Lambda$ is the scale of short-range processes 
that have been integrated out.
Since ChPT is a non-renormalizable theory, 
one must introduce,
for each given order $\nu$, 
new counter-terms 
containing unknown coupling constants,
called the low energy constants (LECs). 
These LECs reflect short-distance physics
that has been integrated out. 
After renormalization at each chiral order, 
the finite part of the LECs 
are to be determined from empirical data, 
see {\it e.g.},~Refs.\cite{fms, fm}. 
Once the LECs are determined, 
${\cal L}_{\rm eff}$ can be used
to make predictions on those observables 
which have not been used as input. 

The present work is concerned with
the renormalization group properties
of the LECs in the pion-nucleon sector.
Our work is motivated by a similar study 
for the pion sector 
by Atance and Schrempp~\cite{as00}.
In Ref.~\cite{as00} it was discussed
that there exist non-trivial infrared fixed points 
in the renormalization group equations 
(RGEs) for the ratios of LECs in the pion sector.  
These fixed points exist 
in the limit $\mu\rightarrow 0$,
where $\mu$ is a renormalization 
scale\footnote{Whether or not 
the fixed points in the limit
$\mu\rightarrow 0$ govern low energy physics 
of the system in question 
is a subtle question, 
reflecting the arbitrariness
of the renormalization scale $\mu$ in ChPT. 
This is in contrast to the case of  
renormalizable theory such as QCD,
where one must choose $\mu$ as a typical energy 
of the process in question
to keep perturbation theory meaningful, 
see {\it e.g.}~Ref.\cite{wein} for details.}. 
Atance and Schrempp estimated,
for the $\nu$=$4$ and $\nu=6$ mesonic chiral Lagrangians,
the LEC ratios 
at the relevant infrared fixed points.
The resulting values of the ratios were found to be
in reasonable agreement with those determined 
from data: 
qualitative agreement in the numerical values
and perfect agreement in the signs.
We describe here an application of 
a similar fixed-point analysis 
to the pion-nucleon Lagrangian. 
We show that, by studying the  infrared fixed points
in the RGEs for the ratios of LECs,
one can estimate the LEC ratios pertaining 
to the pion-nucleon sector
and that the results are in semi-quantitative agreement
with those obtained from direct fitting 
to the experimental data. 
We argue that this agreement can be understood 
by invoking a simple dimensional analysis, and
we discuss our results in the context 
of the Wilsonian renormalization scheme.
 The analysis done in this work is related to the 
method of coupling constant reduction~\cite{oehme, kubo}
 in the sense that both generate the renormalization group invariant 
relations between coupling parameters
(LECs in the present case).\footnote{We thank 
the referee for pointing this out to us.}

\section{RGE fixed points and ratios of LECs}

ChPT is based on a low-energy effective Lagrangian, 
${\cal{L}}_{eff}$,
which is the most general possible Lagrangian
consistent with spontaneously broken chiral symmetry
of QCD.  
${\cal{L}}_{eff}$ is written 
as an expansion in $(Q/\Lambda)^\nu$ 
(see {\it e.g.} Refs.\cite{fms, fmms, ecker} for details): 
\ba
{\cal L}_{\rm eff} ={\cal L}_{\pi\pi}^{(2)}+ 
{\cal L}_{\pi\pi}^{(4)}+{\cal L}_{\pi N}^{(1)}+
{\cal L}_{\pi N}^{(2)}+ {\cal L}_{\pi N}^{(3)}
+\cdots\ ,
\ea
where the superscript denotes the chiral order $\nu$,
and ``$\cdots$'' stands for higher order terms 
in the expansion.
In this work we focus on the third-order 
pion-nucleon Lagrangian,
${\cal L}_{\pi N}^{(3)}$.
Using heavy-baryon chiral perturbation theory, 
and adopting the notational conventions of Ref.\cite{fms},
we write
\ba
{\cal L}_{\pi N}^{(3)}={\cal L}_{\pi N}^{(3), 
{\rm fixed}} +\sum_{i=1}^{23} 
d_i\bar H \tilde O_i H
+\sum_{i=24}^{31}\tilde d_i\bar H \tilde O_i^{\rm div} H \; . 
\label{eq:L3}
\ea 
Here $H$ is the heavy nucleon field,
and $\tilde O_i$ and $\tilde O_i^{\rm div}$ are 
operators constrained by chiral symmetry considerations.
We refer to Ref.\cite{fms} for a complete list of $\tilde O_i$ 
and $\tilde O_i^{\rm div}$.
The LECs, $d_i$ and $\tilde d_i$,
cannot be constrained by symmetry,
and their behaviour is our main concern here.
The ultraviolet (UV) divergences
of loop-diagrams (if they exist) are absorbed in $d_i$ via dimensional regularization
as 
\ba
d_i=d_i^r(\mu) +\frac{(4\pi )^2}{(4\pi F)^2} 
\; \kappa_i \; L (\mu) \; , 
\label{lecri}
\ea
where $\mu$ is a renormalization scale, and 
\ba
L (\mu ) =\frac{\mu^{d-4}}
{(4\pi)^2}\Big(\frac{1}{d-4} - 
\frac{1}{2} [\log 4\pi +1+
\Gamma^\prime (1)]\Big) \; .
\ea 
The constant $F$ is the leading term 
in the quark-mass expansion 
of the pion decay constant $F_\pi$.   
A complete list of analytic expressions for $\kappa_i$ 
can be found in Refs.~\cite{fms,ecker}.
The values of the 
finite renormalized LECs, $d_i^r(\mu)$, 
are to be either fixed by phenomenology or 
estimated with the use of models.

\vspace*{5mm} 

{}From Eq.(\ref{lecri}) one can derive~\cite{ecker} 
a renormalization group equation (RGE)
for the renormalized LEC, $d_i^r (\mu)$,
\ba
\mu\frac{d}{d\mu} d_i^r(\mu)
=-\frac{\kappa_i}
{(4\pi F)^2} \;.
\label{eq:rge}
\ea
 Note that Eq.(\ref{eq:rge}) implies
that there are no fixed points for the renormalized LEC.
We integrate Eq.(\ref{eq:rge}) to obtain 
\ba
d_i^r(\mu )=d_i^r (\mu_0)-\frac{\kappa_i}
{(4\pi F)^2}\log 
\frac{\mu}{\mu_0} \; .  
\label{solr} 
\ea
Eq.(\ref{eq:rge}) leads to the
 RGE for the ratio $d_i^r/d_j^r$ Ref.\cite{as00} ,  
\ba
\mu\frac{d}{d\mu}\left (\frac{ d_i^r(\mu)}
{ d_j^r(\mu)}\right )=
\frac{1}{(4\pi F)^2}\frac{\kappa_j}{d_j^r (\mu)}
\left ( \frac{d_i^r(\mu)}{d_j^r (\mu)}
-\frac{\kappa_i}{\kappa_j}\right ) \; . 
\label{eq:ratio}
\ea 
We emphasize that
these fixed points exist only for the ratios 
of LECs and therefore it might be better
to refer to them as  ``infrared quasi-fixed
points''\cite{LR} or 
as ``Pendleton-Ross infrared fixed points.''\cite{kubo}
However, for the sake of simplicity, we shall continue to use 
the word ``infrared fixed points''.
We note that, 
as can be easily seen from Eq.(\ref{solr}), 
$\kappa_j/d_j^r (\mu)$ in Eq.(\ref{eq:ratio})
is positive 
in the limit $\mu\rightarrow 0$. 
Therefore Eq.(\ref{eq:ratio}) gives 
a non-trivial stable fixed point (f.p.) 
 in the variable $d_i^r/d_j^r$
in the limit 
$\mu\rightarrow 0$\footnote{Eq.(\ref{eq:ratio}) 
has a non-trivial stable f.p. 
in the limit $\mu\rightarrow \infty$ as well. 
However, we will focus on the f.p.'s 
in the limit $\mu\rightarrow 0$  
for a reason to be discussed in Section 3. }, 
\ba
\frac{d_i^r}{d_j^r}\mid_{\rm{f.p.}}=
\frac{\kappa_i}{\kappa_j}~~{\rm for}~ \kappa_i,~\kappa_j\neq 0 \; . 
\label{fp}
\ea 
We note that the derivation of this equation 
involves only few assumptions.
We shall study the consequences 
of Eq.(\ref{fp}) in what follows.
Since $\kappa_i$'s in the pion-nucleon sector  
are known, we can predict the l.h.s. of 
Eq.(\ref{fp}).
Meanwhile, the value of the l.h.s.
can also be obtained from the LECs 
that have been deduced directly
from the experimental data~\cite{fms,fm}.
The comparison of these two determinations
will check the validity of Eq.(\ref{fp}).

For this comparison,
it is useful to consider
the scale-independent LEC,
$\bar d_i$, defined by 
\ba
d_i=\bar d_i + \frac{\kappa_i}{(4\pi F)^2}
\left( (4\pi)^2 L(\mu ) 
-\log \frac{\mu}{M}\right) , 
\label{lecsi}
\ea
where $M$ is the leading term 
in the quark mass expansion 
of the pion mass.
The $\bar d_i$'s have been determined 
phenomenologically in Ref.~\cite{fms}.
Eqs.(\ref{lecri}) and (\ref{lecsi})  
lead to a relation between $\bar d_i$ and 
$d_i^r(\mu)$:
\ba
\bar d_i =d_i^r(\mu) + 
\frac{\kappa_i}{(4\pi F)^2} \log 
\frac{\mu}{M} \;, 
\label{dbr}
\ea 
which implies
\ba
d_i^r(M)=\bar d_i.
\ea
Since the pion mass $M$ is a small scale in ChPT, 
we presume that 
the ratio of  
LECs at $\mu = M$, 
$d_i^r(M)/d_j^r (M)$, is close to its fixed-point value:
\ba
\frac{d_i^r}{d_j^r}\mid_{\rm{f.p.}}\approx d_i^r(M)/d_j^r (M)\ .
\ea
{}From this equation and  Eq.(\ref{fp}), we arrive at our key relation
\ba
\kappa_i /\kappa_j\approx \bar d_i /\bar d_j\ .
\label{comparison}
\ea 
{}This relation allows us to constrain
the ratios of LECs 
without any input from experiments.  
However, a cautionary remark may be in order here.
The present analysis is not applicable to 
a case
in which the $\beta$ function of an LEC vanishes, 
because in such a case
the RGEs for the ratios of LECs involving the LEC 
with the vanishing $\beta$ function do not 
show any non-trivial fixed-point structure
like the one in Eq. (\ref{fp}). 
In general, this could limit the applicability 
of a method described here.
 But, in a particular case of 
the third-order pion-nucleon chiral Lagrangian ${\mathcal L}_{\pi N}^{(3)}$,
all the $\beta$ functions of the LECs as listed 
in Ref.~\cite{ecker} 
are non-zero. In Ref.~\cite{fms}, the authors use a basis different
from Ref.~\cite{ecker}  for the chiral Lagrangian ${\mathcal L}_{\pi N}^{(3)}$
and they list most of the $\beta$ functions for the LECs
 in that basis. We can see that the $\beta$ functions listed in
Ref.\cite{fms} are non-zero.
Therefore,
our analysis is in practice free from the problem of
a vanishing $\beta$ function.
More discussion about the relation in Eq.(\ref{comparison})
will be given later in the text.  
In the following, we compare 
the left- and right-hand sides 
of Eq. (\ref{comparison}).

Experimental information available 
for this comparison may be summarized as follows.
In Refs.\cite{fms,fm}, 
$\bar d_i$'s 
of ${\cal L}_{\pi N}^{(3)}$ 
have been determined 
by analyzing pion-nucleon scattering.
Table 1 gives the values 
of $\bar d_i=d_i^r(\mu\!=\!M)$ determined 
by Fettes {\it et al.}~\cite{fms,fm}. 
In Ref.~\cite{fms}, 
the authors investigated pion-nucleon scattering
in a $\nu$=$3$ ChPT calculation
assuming exact isospin symmetry. 
A more general ChPT $\nu$=$3$ 
analysis of pion-nucleon scattering
was implemented in Ref.~\cite{fm}, 
wherein electromagnetic corrections 
as well as $\nu=2$ isospin violating terms 
of ${\cal L}_{\pi N}^{(2)}$ were taken into account. 
 Since the terms in ${\cal L}_{\pi N}^{(3)}$ 
are higher order correction to 
${\cal L}_{\pi N}^{(2)}$, one 
naturally expects that the determination of 
the counter-terms in ${\cal L}_{\pi N}^{(3)}$
is strongly  influenced by the improvement
in the treatment of ${\cal L}_{\pi N}^{(2)}$. 
Indeed,  Table 1 shows 
drastic differences between the values of the LECs  
obtained in Ref.~\cite{fm} and Ref.~\cite{fms}.
In the following we shall therefore be primarily concerned
with the $\nu=3$ LECs obtained in 
the more general analysis in Ref.~\cite{fm}.  
%

%%%%%%%%%%%%%%% Table 1 %%%%%%%%%%%%%%%%%%
\vskip 0.3cm
  \begin{center}
  Table 1 : \parbox[t]{5.3in}{Phenomenologically
  determined values of the LECs (in ${\rm GeV}^{-2}$) 
 in ${\cal L}_{\pi N}^{(3)}$.
%at the energy scale $\mu=M$.  
The values in the rows labelled 
EXP(Fit 1), EXP(Fit 2) and EXP(Fit 3) are 
taken from Ref.~\cite{fms}, 
while those in the row EXP(2001) are taken from
Ref.~\cite{fm}.}
  \end{center}
  $$
  \begin{array}{|r||r|r|r|r|r|}
  \hline 
 &{\bar d_1} + {\bar d_2} & \bar d_3 & \bar d_5 &\bar d_{14}  
  -\bar d_{15}& \bar d_{18} \\
  \hline
  \hline
 {\rm EXP( Fit}~ 1) & 3.06\pm 0.21 &-3.27\pm 0.73 & 
 0.45\pm 0.42 &-5.65\pm 0.41 
& -1.40 \pm 0.24    \\ \hline
{\rm EXP( Fit}~ 2) & 3.31\pm 0.14&-2.75\pm 0.18
& -0.48\pm 0.06&-5.69\pm 0.28& -
0.78\pm 0.27    \\ \hline
{\rm EXP( Fit}~ 3) & 2.68\pm 0.15&-3.11\pm 0.79
& 0.43\pm 0.49&-5.74\pm 0.29& -
0.83\pm 0.06    \\ \hline
{\rm EXP( 2001 )} & -2.24\pm 0.16 &0.81\pm 0.16  
&0.67\pm 0.11 & -0.63\pm 0.75 & 
-10.14\pm 0.45   \\ \hline
  \end{array}
  $$
  \vskip 0.3cm

%%%%%%%%%%%%%% End of Table 1 %%%%%%%%%%%% 
In addition to the LECs presented in Table 1, 
Eq.(\ref{eq:L3}) contains one more $\nu$=$3$ LEC 
determined from experiment.
In Ref.~\cite{bkm}, $d_6^r(\mu)$ was deduced 
from the isovector charge radius at 
$\mu=\stackrel{\circ}{m}$,
where $\stackrel{\circ}{m}$ is the value of the 
nucleon mass in the chiral limit;
the result is
$d_6^r(\stackrel{\circ}{m})
=-0.13~ {\rm GeV}^{-2}$.   
For our present purposes,
we need to scale $d_6^r(\mu)$ 
down to $\mu\!=\!M$. 
Carrying out  this rescaling using Eq.(\ref{solr}), 
we obtain $d_6^r(M)=\bar d_6 =-2.96~ {\rm GeV}^{-2}$. 
This value of $d_6^r(M)$ 
will also be considered below. 

\vspace*{3mm}

We now examine the two sides of Eq.(\ref{comparison}).
Using the analytic expressions of $\kappa_i$  
given in Refs.~\cite{fms,ecker} and 
$g_A=1.26$, we obtain 
$\kappa_1 +\kappa_2 =-0.42$, $\kappa_3=  0.18$,
$\kappa_5=0.37$ and $\kappa_6=-1.49$. 
The use of these values of $\kappa_i$
in Eq.(\ref{comparison}) leads to RGE-based predictions
of the LEC ratios at the scale $\mu\!=\!M$.
The results are given in the last column (labelled ``RGE")
in Table 2.
This table also shows the LEC ratios determined from  
the experimental information summarized in Table 1.
(For simplicity and because of 
the semi-quantitative nature
of the present study, 
we do not quote errors in the experimental data.)  
Since $\bar d_{15}$ and $\bar d_{18}$ 
are finite LECs and independent of $\mu$~\cite{fms},  
the ratios involving these LECs cannot be determined
from our analysis based on infrared fixed points. 

%%%%%%%%%%%%%%% Table 2 %%%%%%%%%%%%%%%%%%

\vskip 0.3cm
  \begin{center}
  Table 2 : \parbox[t]{5.3in}{The ratios of the
  LECs  in ${\cal L}_{\pi N}^{(3)}$.
  The data in the columns 
  labeled EXP(Fit 1), EXP(Fit 2) and  EXP(Fit 3)
  are taken from Ref.~\cite{fms}, 
 which ignored the isospin breaking term in 
${\cal L}_{\pi N}^{(2)}$. 
The results in the column labeled EXP(2001) 
were obtained with the use of the LECs 
in the last row in Table 1. 
These LECs were determined in Ref.~\cite{fm} 
in an analysis that includes all the relevant terms 
in ${\cal L}_{\pi N}^{(2)}$}. 

  \end{center}
  $$
  \begin{array}{|r||r|r|r|r|r|}
  \hline
 {\rm Ratio~ of ~LEC}s &{\rm EXP}~({\rm Fit}~1) 
 &{\rm EXP}~({\rm Fit}~2)& {\rm EXP}~({\rm Fit}~3)&
{\rm EXP}~ (2001) &{\rm  RGE} \\
  \hline
  \hline
  (\bar d_1+\bar d_2)/\bar d_3 &-0.94& -1.23&-0.86 &-2.77 & -2.33   \\ 
\hline

(\bar d_1+\bar d_2)/\bar d_5 &6.8& -6.9&6.23 &-3.34&  -1.14  \\ \hline
{\bar d_3/\bar d_5} &-7.27& 5.79& -7.23& 1.21& 0.49  \\ \hline
(\bar d_1+\bar d_2)/\bar d_6 &-1.03& -1.12&-0.91 &0.76& 0.28  \\ \hline
\bar d_3/\bar d_6 &1.1& 0.93&1.1 &-0.27& -0.12   \\ \hline
\bar d_5/\bar d_6 &-0.15& 0.16&-0.15 &-0.23& -0.25   \\ \hline
  \end{array}
  $$
  \vskip 0.3cm
%%%%%%%%%%%%%% End of Table 2 %%%%%%%%%%%%

In Table 2 we note 
that the ratios given in the column
labeled EXP (2001) and those obtained
in our RGE analysis
show semi-quantitative agreement; 
the signs are all in agreement
and the magnitudes exhibit 
a similar general tendency.
We note that the results of our RGE analysis
show definite disagreement with
those of the earlier (presumably less reliable)
empirical determinations of the LECs.   
This may be taken as an indication
that we can profitably use 
a fixed-point analysis like the one described here  
as a constraint in determining
the LECs from data.
 To illustrate this point,
we present here predictions on the ratios of
LECs that involve LECs which 
have not so far been determined by experiments.
Although the number of ratios that can be predicted
is in fact quite large, we consider here only  some
examples. Predictions for these selected cases
are given in Table 3.

%%%%%%%%%%%%%% Table 3 %%%%%%%%%%%%%%%%%%

\vskip 0.3cm
  \begin{center}
  Table 3: \parbox[t]{5.3in}
  { Examples of the predicted ratios of the
  LECs that appear in ${\cal L}_{\pi N}^{(3)}$~\cite{fms}.
  The symbol $d_i^{(j)}$ is defined
  as $d_i^{(j)}\equiv\bar d_i/\bar d_j$.}

  \end{center}
  $$
  \begin{array}{|r||r|r||r|r||r|r||r|}
  \hline
 d_8^{(5)}& 1.35&d_{10}^{(5)}&0.63 &d_8^{(6)}& -0.34&d_{10}^{(6)}&-0.16\\
\hline
d_{11}^{(5)}&-0.29&d_{12}^{(5)}&-2.33 &d_{11}^{(6)}& 0.07&d_{12}^{(6)}&0.58\\
\hline
d_{13}^{(5)}& -0.72&d_{14}^{(5)}&-1.7&d_{13}^{(6)}& 0.18&d_{14}^{(6)}&0.42 \\
\hline
d_{16}^{(5)}&
1.03&d_{20}^{(5)}&3.41 &d_{16}^{(6)}& -0.26&d_{20}^{(6)}&-0.85  \\
\hline
 \end{array}
  $$
  \vskip 0.3cm
%%%%%%%%%%%%%% End of Table 3 %%%%%%%%%%%%

\section{Discussion and summary}

We observed that the RGE for the ratio 
$d_i^r/d_j^r$ of the renormalized
LEC has a non-trivial  infrared fixed point given 
by  Eq.(\ref{fp}). We now
discuss the implication 
of Eq.(\ref{fp}) in the context of 
a simple dimensional analysis~\cite{georgi,manohar}.
To this end, we first briefly review the naive dimensional
analysis given in Ref.~\cite{georgi}.
Consider the $\pi$-$\pi$ scattering amplitude 
at order $Q^4$.
The amplitude of a one-loop diagram involving 
the lowest order Lagrangian 
${\cal L}^{(2)}_{\pi \pi}$ is given by
\ba
 \frac{Q^4}{F^4}\frac{1}{(4\pi)^2}\log\mu 
 + \cdots \; . 
\label{eq:pipiloop}
\ea
{}The necessary counter-terms come from 
the $\nu$=$4$ Lagrangian ${\cal L}^{(4)}_{\pi \pi}$. 
A term like
\ba
\frac{F^2}{\mu_{\chi SB}^2} 
tr (\partial^\mu\Sigma\partial^\nu\Sigma
\partial_\mu\Sigma^\dagger\partial_\nu\Sigma^\dagger)
\ea
gives a scattering amplitude of order $cQ^4/F^4$, 
where $c$ is given by 
$c=F^2/\mu_{\chi SB}^2$.
Since the total $\pi$-$\pi$ scattering amplitude 
should not depend on the renormalization 
scale $\mu$, 
a shift in $\mu$ should be 
compensated by a shift in $c$. 
Therefore a change in $\mu$ of order one 
produces a change in $c$ of 
order $\delta c\sim 1/(4\pi )^2$. 
Then, barring the accidental fine tuning 
of the parameters, $c$ must be of the 
order of $\delta c$,
\ba
c\sim \delta c\sim \frac{1}{(4\pi)^2}.
\ea
This implies that $\mu_{\chi SB}\sim 4\pi F $~\cite{georgi}. 
Now, an advantage of using the RGE 
is that one can calculate $\delta c$ explicitly. 
To show this point in our case, 
we return to Eq.(\ref{solr}).  
As stated, observables should be independent 
of $\mu$.
{}From Eq.(\ref{solr}) we can see 
that a change in $\mu$ 
produces a change in the LEC $d_i^r (\mu)$: 
\ba
\delta d_i^r&\equiv& d_i^r 
(\mu )-d_i^r (\mu_0)\no
&=& - \frac{ \kappa_i}{(4\pi F)^2} 
\log\frac{\mu}{\mu_0} \; . 
\label{eq:deltaLEC}
\ea
As discussed above, from
$\delta d_i^r(\mu ) \sim d_i^r(\mu )$, 
we may infer
\ba
\frac{d_i^r (\mu)}{d_j^r 
(\mu)}\sim \frac{\kappa_i}{\kappa_j}.
\ea 
We now discuss, 
in the context of a Wilsonian renormalization
group, whether the fixed points 
featuring in Eq.(\ref{fp}) are relevant to
low energy physics.\footnote{We thank M. Rho for 
suggesting this viewpoint.}
To do this, we follow Ref.\cite{hy00}
and derive a Wilsonian RGE
that includes quadratic divergences as well as 
logarithmic divergences. 
To preserve chiral symmetry, 
dimensional regularization is adopted,
and the quadratic divergences
are identified by the following replacement 
\cite{hy00, velt} :
\ba
\int \frac{d^nk}{i(2\pi)^n}\frac{1}{-k^2}
\rightarrow\frac{\Lambda^2}{(4\pi 
)^2}~,~~~
\int \frac{d^nk}{i(2\pi)^n}\frac{k_\mu k\nu}
{[-k^2]^2}\rightarrow -\frac{\Lambda^2}{2(4\pi )^2}g_{\mu\nu}\; ,
\label{c1hy00}
\ea
where the scale $\Lambda$ has the meaning of a naive cutoff.
These replacements are known to preserve 
chiral symmetry at the 
one-loop order \cite{harada}. 
For illustration,
we again use the one-loop diagram 
of the $\pi$-$\pi$ scattering. 
Since the logarithmic divergences in 
the  $\pi$-$\pi$ scattering are
well known, we focus on the quadratic divergences.
Let us take a one-loop diagram
 with one vertex from ${\cal L}_{\pi\pi}^{(2)}$
and the other one from  ${\cal L}_{\pi\pi}^{(4)}$.
The vertex from  ${\cal L}_{\pi\pi}^{(2)}$ is of the form 
 $p^2/F^2$ and the one from  ${\cal L}_{\pi\pi}^{(4)}$ 
 has the form $L_i~ p^4/F^4$, 
where $L_i$ is a LEC 
 in ${\cal L}_{\pi\pi}^{(4)}$. 
We consider the contribution of a particular term in which  
 only two of the six derivatives act on 
the internal lines. 
{}If we apply Eq.(\ref{c1hy00}) to this term, 
its quadratic divergence is given schematically by
\ba
L_i~ \frac{p^4}{F^6}\int \frac{d^nk}{(2\pi)^n}\frac{1}{k^2}
\rightarrow L_i~\frac{p^4}{F^4}\frac{\Lambda^2}{(4\pi F)^2}\; .
\label{qdex}
\ea
We may assume that there are similar quadratic divergences 
in the $\pi$-$N$ sector.  We can then generalize the RGE
in Eq. (\ref{eq:rge}) and write: 
\ba
\mu\frac{d}{d\mu} d_i^r(\mu)=-\frac{\kappa_i}{\Lambda_{\chi SB}^2} 
+ g_i\frac{\mu^2}{\Lambda_{\chi SB}^2}d_i^r (\mu)
+h_i\frac{\mu^2}
{\Lambda_{\chi SB}^2}~,\label{grg}
\ea
where $\Lambda_{\chi SB}=4\pi F$ 
and $g_i$ and $h_i$ are calculable
constants.
The term with $g_i$ in Eq. (\ref{grg}) is similar to the quadratic
divergence in Eq. (\ref{qdex}),
and the term with $h_i$ is a possible
quadratic divergence with all vertices 
coming from ${\cal L}_{\pi N}^{(1)}$
and ${\cal L}_{\pi N}^{(2)}$.
{}It is easy to obtain from Eq. (\ref{grg})
a modified RGE for the ratio of the LECs
\ba
\mu\frac{d}{d\mu}\left(\frac{ d_i^r(\mu)}
{ d_j^r (\mu)}\right )
 &=&\frac{\kappa_j}{d_j^r (\mu)}
[\frac{1}{\Lambda_{\chi SB}^2} 
(\frac{d_i^r(\mu)}{d_j^r (\mu)}
-\frac{\kappa_i}{\kappa_j})\no
 &&~~~+\frac{1}{\kappa_j}
\frac{\mu^2}{\Lambda_{\chi SB}^2}
( g_id_i^r-g_jd_i^r +h_i-h_j\frac{ d_i^r(\mu)}
{ d_j^r (\mu)} ) ].
\label{WRGE}
\ea
In this Wilsonian RGE, 
the terms with $\mu^2/\Lambda_{\chi SB}^2$
becomes negligible in a low-energy regime (${\mu^2}\ll 
{\Lambda_{\chi SB}^2}$), and, as a result, we recover the
RGE in Eq.(\ref{eq:ratio}).
This feature suggests that the infrared fixed points in 
Eq.(\ref{fp}) governs low
energy physics qualitatively up to
$\mu^2/\Lambda_{\chi SB}^2$ corrections.

Finally we discuss the significance of 
our infrared fixed points in relation to QCD.
As is well known, 
the form of a ChPT Lagrangian is chosen 
using the chiral symmetry properties of QCD 
without invoking detailed dynamical information on QCD.
This feature, combined with the fact 
that a fixed point (if any) 
of a given field theory is generally 
independent of the initial conditions,
may make one wonder whether there is any physical
relation between QCD and the infrared fixed points
considered in the present work. 
Although a formal treatment of this issue 
goes beyond the scope of this article,
we present a plausibility argument that indicates
that these fixed points are likely to 
have physical significance.
First, we invoke the general tenet of ChPT that,
if the LECs appearing in ChPT
are determined to reproduce the 
relevant low-energy observables,
these LECs effectively subsume 
the short-distance physics of QCD. 
The phenomenological success of ChPT
indicates the basic soundness of this tenet;
in this phenomenological sense
ChPT is physically related to QCD.
Meanwhile, once the ChPT Lagrangian is specified,
its renormalization group properties and fixed points
are automatic consequences of the 
structure of the given Lagrangian.  
These two aspects lead us to expect that   
the infrared fixed points considered in the present work
are physically (if not formally) related to QCD.
Admittedly, this is just a plausibility argument.
In this connection, there exists a very illuminating 
study by Harada and Yamawaki~\cite{HY01}.
These authors investigated a hidden local symmetry (HLS)
approach
as an effective theory of QCD, by matching HLS and QCD
at a matching scale $\Lambda \sim 1 {\rm GeV}$.
They found that in HLS there occur three fixed points
and one fixed line in the physical region but that
only one of them, a so-called vector manifestation (VM) point,
has correspondence with QCD. 
 We recognize that 
it is not possible to make a {\it direct} connection
between the results reported in Ref.~\cite{HY01}
and those obtained in our present study,
because Ref.~\cite{HY01} discusses
the fixed points in the LECs themselves
whereas we are considering here the ratios of LECs.
After stating this warning,
we may add the following speculative remark. In
our consideration of the RGE for ChPT,
we have found only one fixed point for each ratio of the LECs.
It therefore seems natural to expect that
the fixed points determined in the present work have
bearing upon QCD.
Again, this is just an intuitive argument, but we believe 
that, with this caveat kept in mind, the results
given in this work can be of practical use.

\vspace*{5mm}
In summary,  we have estimated the ratios
of the scale invariant LECs, $\bar d_i/\bar d_j$,
by studying the fixed point structure of the RGE 
for the ratios of the renormalized LECs.
We have found that the ratios of the LECs 
determined from experiments~\cite{fm,bkm}
and the ones estimated by our RGE analysis 
agree semi-quantitatively.  
We have given a plausible explanation 
for this agreement,
invoking a simple dimensional analysis 
and the Wilsonian RGE. 
To address the subtle problem in ChPT, 
whether or not the limit $\mu\rightarrow 0$ 
can be associated with a low-energy scale
of the system, 
we have resorted to the Wilsonian RGE, Eq.(\ref{WRGE}),
and argued that the fixed points in Eq.(\ref{fp})
govern low-energy physics qualitatively up to
$\mu^2/\Lambda_{\chi SB}^2$ corrections. 
This argument combined with the fact that 
the pion mass $M$ is a small scale in ChPT leads us to presume that 
the ratio of the LECs at $\mu = M$, 
$d_i^r(M)/d_j^r (M)$, is close to its infrared fixed-point value.
We should perhaps repeat here our cautionary remark (see sect. 2)
that what we call ``infrared fixed point'' should, to be precise, 
 be referred to as 
``infrared quasi-fixed
point''\cite{LR} or as
 ``Pendleton-Ross infrared fixed point.''\cite{kubo}.
It is hoped that
an analysis like the one described here
may be useful in placing constraints
on the values of other LECs in ChPT as well.

\section*{Acknowledgements}

We thank S. Ando for discussion. Our thanks 
are also due to M. Harada, 
U.-G. Meissner, and
M. Rho for their helpful comments.
This work is supported in part by 
the U.S. National Science Foundation, 
Grant Nos. PHY-0140214 and INT-9730847.
The work of YK is supported in part 
by the Brain Korea 21 Project 
of the Ministry of Education,
by the KOSEF R01-1999-000-00017-0
and by the KRF 2001-015-DP0085.

\end{document}